# Applications of sinusoidal phase modulation in temporal optics to highlight some properties of the Fourier transform

K Hammani, [1,2] J. Fatome, [1] and C Finot [1,2]

[1] Laboratoire Interdisciplinaire Carnot de Bourgogne, UMR 6303 CNRS - Université de Bourgogne-Franche-Comté, 9 avenue Alain Savary, BP 47870, 21078 Dijon Cedex, France

[2] Département de Physique, Faculté des Sciences et Techniques, Université de Bourgogne, 9 avenue Alain Savary, BP 47870, 21078 Dijon Cedex, France

E-mail : christophe.finot@u-bourgogne.fr

**Abstract.** Fourier analysis plays a major role in the analysis and understanding of many phenomena in physics and contemporary engineering. However, students, who have often discovered this notion through numerical tools, do not necessarily understand all the richness that can be derived from joint analysis in the temporal and spectral domains, particularly in the field of optics. As part of the second year of the Master's degree in Physics Lasers and Materials at the University of Burgundy, we have set up a set of experiments to highlight these concepts and to show, on a non-trivial example of periodic phase modulation, the precautions to be taken in the interpretation of the various experimentally accessible spectra. This hands-on session, made possible through the use of the research infrastructure, is also an introduction to the use of standard optical telecommunications equipment.

## 1. Introduction

The vast majority of optical instruments such as photodiodes, CCD sensors, or even the eye are above all sensitive to the intensity of the light wave. Thus, optical telecommunications were initially based on a binary modulation of the intensity of the transmitted wave [1-3]. This made it possible to neglect to a large extent the effects affecting the phase of the wave which is much more difficult to control. However, technological advances in the last two decades have fundamentally challenged this approach by making coherent communications possible, i.e. communications based on the modulation of the optical carrier phase. Thenceforth, by supplanting traditional On-Off Keying intensity formats, the advent of coherent communications marks a major turning point in broadband technologies at the dawn of the 21st century [4, 5].

In this contribution, we describe a lab session that aimed to highlight various signatures of the simplest phase modulation that is possible, i.e. a sinusoidal temporal phase modulation of an initial continuous wave. The temporal waveform of a signal as well as its spectrum (i.e., frequency content) are two fundamental vehicles enabling to characterize a signal. Following the pioneering work of the French physicist Joseph Fourier [6] who was born in Burgundy 250 years ago, whenever an operation is performed on the waveform of a signal in the time domain (real space), a corresponding modification is applied to the spectrum of the signal in the frequency domain (reciprocal space), and vice versa. Therefore, we want to form an idea of the corollary that the application of a phase modulation in the time domain deeply affects the signal in the frequency domain. We wish here to highlight the possibilities offered by Fourier analysis [7, 8], which plays a major role in countless branches of physics and engineering [9] such as electronics, acoustics, as well as optics [10-13]. While master students are used to manipulating this concept in the case of real initial signals (for example, in the case of Fraunhofer diffraction caused by an aperture), we found that they had much more difficulties in managing phase-modulated signals. Moreover, throughout this lab session, we wanted to avoid an exclusive numerical approach and we invited our students to manipulate on state-of-the-art equipments of a genuine research platform where they can benefit from a large set of instruments to directly monitor different consequences of the temporal phase modulation.

The present report is organized as follows. We will first present the laboratory equipment made available to our master students. We then summarize the series of experiments they had to conduct as well as the key points of each manipulation. Before concluding, we will also discuss how we have evaluated the work of our students that follow a master's degree programme specializing in Physics, Laser and Materials (French Physics master's degree taught at the University of Burgundy between 2008 and 2016).

## 2. Experimental setup

*2.1. Research environnement*

The proposed lab session is based on the PICASSO platform (Innovation and Design Platform for the Analysis and Simulation of Optical Systems) at the University of Burgundy (UB). More precisely, this platform, part of the Solitons, Lasers and Optical Communications team of the CARNOT Interdisciplinary Laboratory of Burgundy, is specialized in high-speed telecommunications and benefits from state-of-the-art lightwave characterization equipment. Essentially dedicated to research studies, we planned here for the first time to involve these facilities as part of a pedagogic activity aimed at a Master's level audience

(excluding internships in the laboratory). This point was positively noted by our students who highly appreciated using high-end equipment with a higher richness than the devices usually devoted to pedagogy. The interface and the technical documentation of the equipment exclusively in English language did not seem to discourage the French students who quickly gained a sufficient level of autonomy for the most basic operations. It should be noted that the technical datasheets summarizing the main characteristics of each device were included into a booklet given to the students well in advance (in this article, we provide the detailed references of all devices so that the reader could easily find the corresponding datasheet using any internet search engine). In the same way, we made the students aware of the cost of acquiring such equipment and the positioning of the equipment at the commercial level (mid-range or high-end).

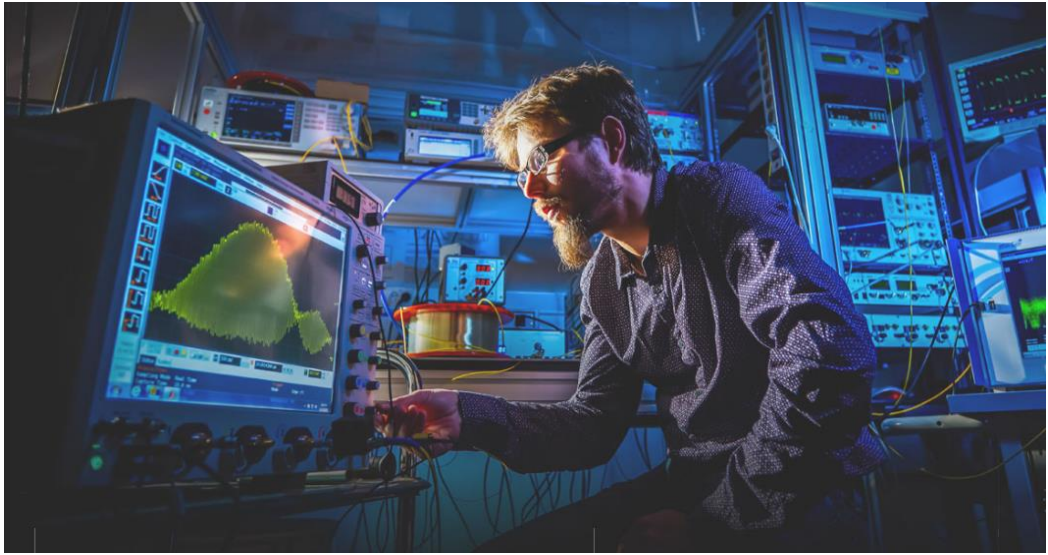

**Figure 1** – View of one part of the experimental testbed dedicated to the characterization of optical components for high speed signal generation and transmission. The students spent three hours running different experiments on this testbed platform. In the fore front of the picture is Dr. Julien Fatome. Photo by Alexis Chezière.

## 2.2. Equipment made available to the students

### 2.2.1. Optical source and modulation

The initial optical wave studied in the framework of this lab session consists in a classical optical communication signal. A laser diode (ECL technology: external cavity laser diode, model OSICS ECL-1560 manufactured by Yenista) delivers a quasi-monochromatic light beam at a wavelength within the conventional band of optical transmissions. The optical complex field of this initial continuous wave $\psi_c(t)$ can therefore be written as $\psi_c(t) = \psi_0 \exp(i\,\omega_0\,t)$ where $\psi_0$ is the initial amplitude of the wave and $\omega_0$ its carrier frequency, directly related to its wavelength $\lambda$ by $\lambda = 2\pi c/\omega_0$ ($c$ being the velocity of light in vacuum). This signal centered on 1550 nm (i.e. $\omega_0 = 1\ 216$ rad.s$^{-1}$) is then phase-modulated using a Lithium Niobate modulator (LiNbO$_3$, sold by Thorlabs, 10 GHz Phase Modulators, model LN65S-FC). Thanks to the Pockels electro-optic effect, this device converts an incoming electrical modulation into a temporal modulation of the optical phase of the wave [3]. The modulator is characterized by an electrical voltage $V_\pi$ of typically 6 Volts, which corresponds to the external voltage that must be applied to the modulator to induce an optical phase shift of π. In our setup, the phase modulator is directly powered by an electrical clock delivering a perfectly periodic sinusoidal signal at a frequency of 10 GHz (Anritsu RF/Microwave Signal Generator MG3692B) with an output average power up to 23 dBm. After phase modulation, the optical field becomes $\psi_m(t) = \psi_c \exp[i\,A_m \cos(\omega_m t)]$, where $\omega_m$ is the angular frequency of the sinusoidal modulation and $A_m$ is the peak deviation of the phase modulation. The temporal phase change $\phi(t)$ induces a change of the instantaneous frequency $\delta\omega(t)$ given by the time derivative of the phase profile : $\delta\omega(t) = -d\phi(t)/dt$. The use of equipment typical of optical telecommunications makes it easier to handle frequencies in the GHz range, which, as we will see later, is an essential condition for observing an optical spectrum with a sufficient accuracy. It should be also noted that all the optical elements used are provided with optical connectors so that no optical alignment is required: the lab session can therefore focus on physical discussions and will not be slowed down by any problems of fine tuning to be carried out. Finally, the use of a ligth source and modulator with polarization maintaining elements ensures an increased stability and a very low impact of surrounding conditions.

### 2.2.2. Temporal analysis

It is rather cumbersome to measure the complete temporal profile of the optical field $\psi(t)$ directly. Very specific experimental methods such as the Frequency Resolved Optical Gating approach [14] exist but their use remains quite complex and far exceeds the objectives of the course. Visualizing the temporal information contained in the phase requires approaches of an interferometric nature, such as the Michelson or Mach-Zehnder devices. For this lab session, we didn't want to involve such devices. Thus, we focused on the detection of the temporal intensity profile alone $I(t) \propto |\psi(t)|^2$. We have a high bandwidth photodiode (>50GHz, from u2t photonics) connected to an oscilloscope with a suitable electrical analysis bandwidth (33 GHz, Agilent Infinium, DXO-X-93304Q). As the signal under study is periodic, an optical sampling oscilloscope with a bandwidth of 1 THz is also available in the laboratory (EXFO, PSO-100). The students also had a power meter that allows them to measure the average power of the light beam. It should be noted that in order to complement this lab session, students had another lab session dedicated specifically to the questions linked to the temporal characterization of optical nano- and picosecond pulses where the constraints on the analysis bandwidths are detailed.

*2.2.3. Spectral analysis*

The main objective of this lab session is to stimulate discussions on the analysis in the reciprocal domain, i.e. the use of the Fourier transform. The expression of the field $\psi(\omega)$ in the spectral domain is thus given by the following Fourier transform:

$$\psi(\omega) = \int_{-\infty}^{\infty} \psi(t)\, e^{i\omega t} dt \ . \tag{1}$$

Similarly to the time domain, it is experimentally difficult to access $\psi(\omega)$ in amplitude and phase. Thus, with conventional devices, only the spectral intensity profile $S(\omega) \propto |\psi|^2$ is easily recordable. This parameter represents the optical spectrum, which is a measure commonly used in the field of spectroscopy, for example. The students had to test two different optical spectrum analyzers (OSA) and to compare the results obtained from both OSA providing two levels of resolutions: 0.07 nm for the first one (mid-range, Anritsu MS9710B) and < 0.1 pm for the other (high-end product, APEX AP2441B).

As far as frequency analysis is concerned, we did not want to limit ourselves to the optical spectrum. Thus, we also studied the electrical spectrum (also known as the radio-frequency spectrum) which corresponds to the Fourier transform of the intensity temporal profile:

$$R(\omega) \propto \left| \int_{-\infty}^{\infty} |\psi(t)^2|\, e^{i\omega t} dt \right| \tag{2}$$

We have provided students with an electrical spectrum analyzer (ESA) with a 26 GHz bandwidth (Agilent EXA Signal Analyzer, N9010A). To benefit from the highest measurement dynamics, we chose a dedicated device but the students also discovered the mathematical functionalities now routinely implemented in the latest generations of real-time oscilloscopes.

*2.2.4. Optical spectral shaping of light*

The last step of the session is to test the impact of a change in the phase or amplitude properties of the optical spectrum. We compared two optical elements. The first one was a simple single-mode optical fiber. In that case, the dispersion of the fiber imprints a spectral parabolic phase that is easily modelled by:

$$\psi_{out}(\omega) = \psi_{in}(\omega) \exp\left(\frac{i\, \beta_2\, L\, \omega^2}{2}\right) \tag{3}$$

with $\beta_2$ the dispersion coefficient of the group velocities and $L$ the length of the fiber. We used a spool of the most standard fiber of the telecom industry, the SMF-28 fiber, those properties are normalized according to the recommendation G.652 of the International Telecommunication Union [15]. The dispersion coefficient of this fiber is $\beta_2$ = -20 $10^{-3}$ ps$^2$/m and its length is 2 km.

To go beyond a simple parabolic change of the spectral phase, we also provided students with a programmable optical shaping device based on a method similar to the 4-f method [16]. Previously restricted to only a few laboratories and femtosecond applications [17], linear optical shaping has now emerged in the wavelength domain of optical telecommunications with now easy-to-use commercial devices [18] (we use a Finisar Waveshaper 1000S device). The shaper is thus able to impose a spectral phase and amplitude transfer function $T(\omega)$ such that $\psi_{out}(\omega) = T(\omega)\,\psi_{in}(\omega)$. In this lab session, we limited ourselves to the use of simple transfer functions directly embedded in the control software such as the addition of a parabolic phase $T(\omega) = \exp(i\,A\,\omega^2)$ (with $A$ being a positive or negative coefficient) or amplitude filtering by a Gaussian shape whose central frequency $\omega_1$ and bandwidth $\omega_2$ can be varied: $T(\omega) = \exp(-(\omega-\omega_1)^2/\omega_2^2)$. Note that in order to compensate for the optical losses induced by the spectral shaper, we can take advantage of the optical gain provided by an erbium doped fiber amplifier that has now become a standard device of the telecom industry [1].

*2.3. Complete setup*

The complete setup is schematized in Figure 2 where we have distinguished through different colors the parts dealing with an electrical signal from those dealing with an optical signal. The practical implementation does not present any particular difficulty, the whole setup being fibered [6]. Thus, it is the students themselves who handled and connected the various devices as the lab session goes on. This lab session was an opportunity to make our students aware of the precautions to be taken when handling optical connectors (prior cleaning, laser risks) or RF electrostatic sensitive devices. In particular, we stressed the importance of impedance matching conditions and electrostatic shocks. In total, the equipment used in this experiment represents an acquisition budget of approximately 500 k€ (mainly for detection equipment, the optical sampling oscilloscope, the real-time electrical oscilloscope and the high-resolution optical spectrum analyzer costing more than 100 k€ each). Such a cost made it crucial to cooperate with a platform dedicated to research, such a collaboration being the only way able to have these elements available for master students.

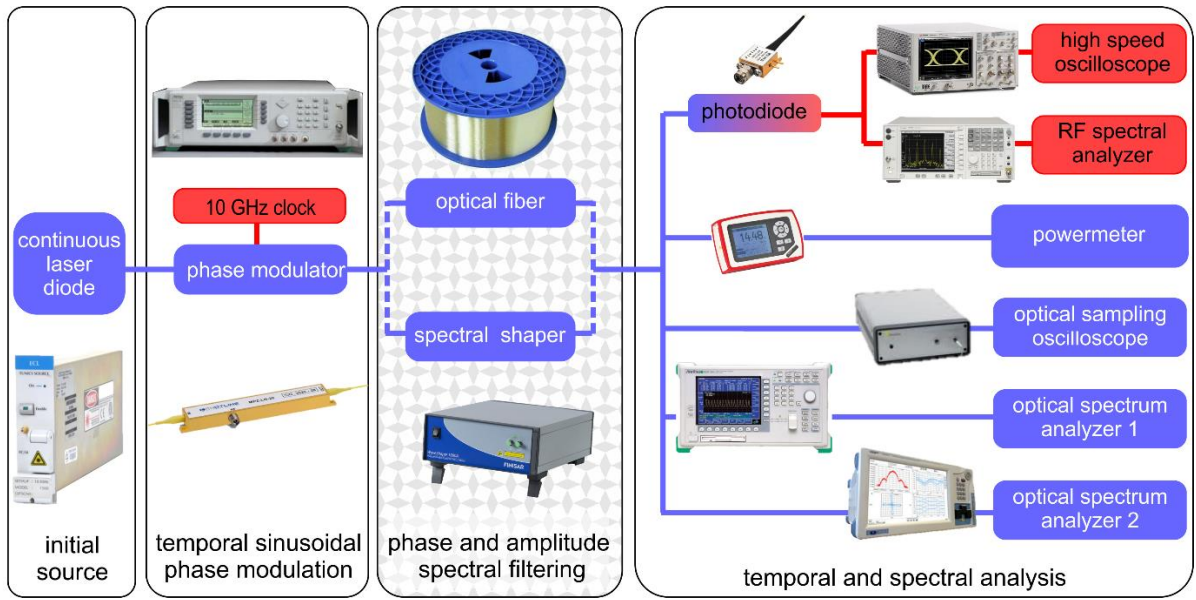

**Figure 2** – Experimental setup. Blue elements are for the devices and connections dealing with an optical wave whereas the red color corresponds to equipment dealing with electrical signals.

### 3. Description of the experimental work

The practical session took place in three subsequent stages, with the complexity of the experiments gradually increasing. The students had first to think over the spectral analysis of a continuous signal before studying the various consequences of the temporal phase modulation. The last step was to manipulate the basic spectral properties of the resulting signal.

*3.1 Analysis of the continuous optical wave*
To begin with this experimental work and to get to grips with the different diagnostic elements, we first studied the spectral properties of the optical signal emitted by the ECL diode. The students had to determine the spectral linewidth at half-maximum of the diode and the noise level of the component. These a priori simple characteristics nevertheless require a good understanding of the settings of the optical spectrum analyzer. It is therefore important to choose the finest resolution. For the Anritsu MS9710B model, the resolution varies between 0.07 and 1 nm. The comparison of the measurements shows that the resolution is crucial for the shape of the recorded signal (Fig. 3a). None of the proposed resolutions (0.07 nm, 0.2 nm and 0.5 nm) allows to conclude on the spectral width of the source, the technical documentation of the laser indicating a value of 150 kHz. With the higher end model (resolution of only 5 MHz), students can infer that the source has a spectral width of less than 5 MHz, which is in accordance with the technical documentation. Therefore, for the highly coherent source we use and given the resolution of the current

OSA, it is not possible to provide a value of the spectral linewidth by a direct measurement in the optical domain.

Regarding the noise level measurement, a good understanding of the technical data is once again necessary. Indeed, depending on the settings used for the instrument and its acquisition mode, the measurements can be limited by the electronic acquisition noise. The students therefore had to think over the nature of the noise floor recorded on the OSA, i.e. the limit of the optoelectronic detection or noise that can be attributed to the tested component.

To complete this first step, we studied the signal recorded by the RF analyzer. The students had to check that in the basic configuration, the RF spectrum did not provide any relevant measurements on the spectral width of the diode. Unlike the optical analyzer, which gave important information about wavelength (i.e. the carrier frequency $\omega_0$), the RF spectrum is only sensitive to the intensity profile of the envelope of the wave. However, let us mention that more advanced heterodyne architectures can provide very accurate access to the diode linewidth.

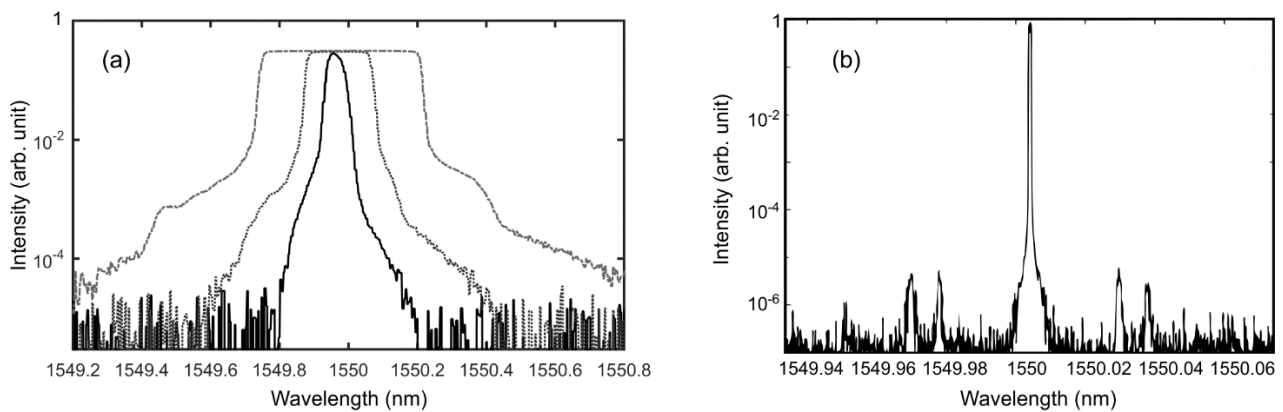

**Figure 3** – Optical spectra of the continuous wave laser. **(a)** Results recorded with different resolution (0.07 nm, 0.2 nm and 0.5 for the solid black, dotted and dashed grey curves respectively) with the Anritsu MS9710B device. **(b)** Results recorded on a high-resolution osa.

*3.2 Spectral analysis of the phase modulated light*

The core of this lab session lies in the spectral analysis of the phase-modulated signal. The students now operate the phase modulator by supplying it with a sinusoidal voltage varying at a frequency of 10 GHz and with an average power of 20 dBm.

3.2.1 In the RF spectral domain.

Students started by observing the RF spectrum of the phase modulated signal. Searching for a signature of the modulation, they increased the sensitivity to the maximum of the ESA and succeeded in detecting an extremely weak line slightly emerging from the noise at the frequency $\omega_m$. Nevertheless, they are surprised to find that when the ECL diode is switched off, this low peak remains: it cannot therefore be characteristic of the optical wave. It is in fact due to the electromagnetic radiation from the generator and the unshielded cables and connectors that are involved in the setup. Optical phase modulation therefore does not impact the RF spectrum, unlike phase modulation as traditionally used in the field of frequency modulation-based radio or microwave communications [19, 20]. In the latter case, the RF spectrum analyzer allows to see the impact of modulation on the carrier frequency and is not restricted, as in the case of optics, to information on the envelope. To confirm that there was no impact on the envelope intensity profile, the students were able to visualize the signal on the fast oscilloscopes at their disposal. They also checked that the presence of the phase modulation was not involving additional power losses on the optical signal (except for the insertion losses of the component).

3.2.2 In the optical spectral domain
They then observed the optical spectrum of the signal. They first used an analyzer with a low resolution (Anritsu device). In contrast to the RF analysis previously carried out, a very significant change in the optical spectrum is observed in the presence of modulation (Fig. 4). Even if the central wavelength does not change ($\omega_0$ is not affected), the resulting spectrum is now very different from an isolated (even widened) peak characteristic of a continuous signal. The spectrum is symmetrical and has a significant width and a complex structure with regularly spaced bumps. By changing the amplitude of the electrical (and therefore optical) modulation, the students verified that the width and shape of the spectrum varies according to $A_m$, while the intensity profile $I(t)$ remains unchanged. This point is an apparent violation of a simplified relationship on which many students base their qualitative thinking: for a Fourier transform-limited waveform, the wider a spectrum, the shorter the time structure. As we can see with this simple example, for the complex light field, this relationship must be handled with care.

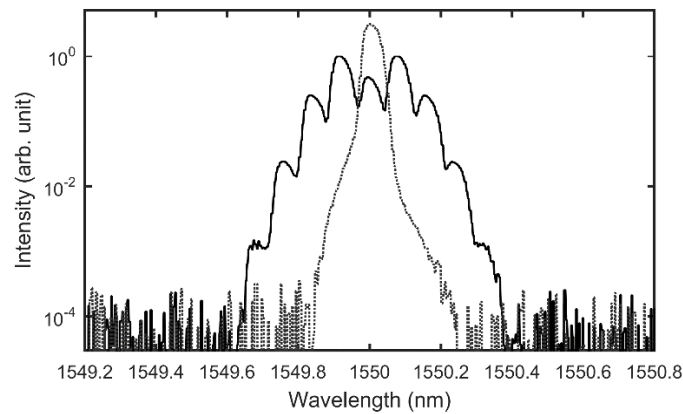

**Figure 4** – Optical spectrum of the phase modulated signal (solid line) compared with the spectrum without phase modulation (grey dotted line). The experimental results are recorded on an OSA with a spectral resolution of 0.07 nm. The phase modulator is driven by a sinusoidal electrical modulation at a frequency of 10 GHz and with an average power of 20 dBm.

The use of an OSA with a higher resolution is essential to reveal the details of this structure and to make the link with analytical theory. Examples of the results obtained with a high-end optical device are shown in Figure 5. First of all, these recordings allow seeing clearly a spectrum of lines that was partially occulted by the limited resolution of the Anritsu device. This comb nature is in perfect agreement with the periodic nature of the signal under test. Consequently, the spectral analysis of the signal becomes limited to the analysis of a set of discrete lines. By setting the OSA display in frequency units rather than in wavelength units, the students were able to check that the frequency spacing between two lines corresponds exactly to the modulation frequency $\omega_m$ of the signal: by modifying $\omega_m$ on the electrical generator, they can straightforwardly reduce or increase the spectral spacing without changing the overall shape of the spectrum. We also note the symmetrical nature of the experimental spectrum, the decreasing amplitude of the components (for the range of $A_m$ under study and excluding the central component) as well as the excellent signal-to-noise ratio of the optical signal. The students have greatly appreciated the great measurement dynamics offered by the OSA device, which are capable of accurate measurements over more than 6 orders of magnitude.

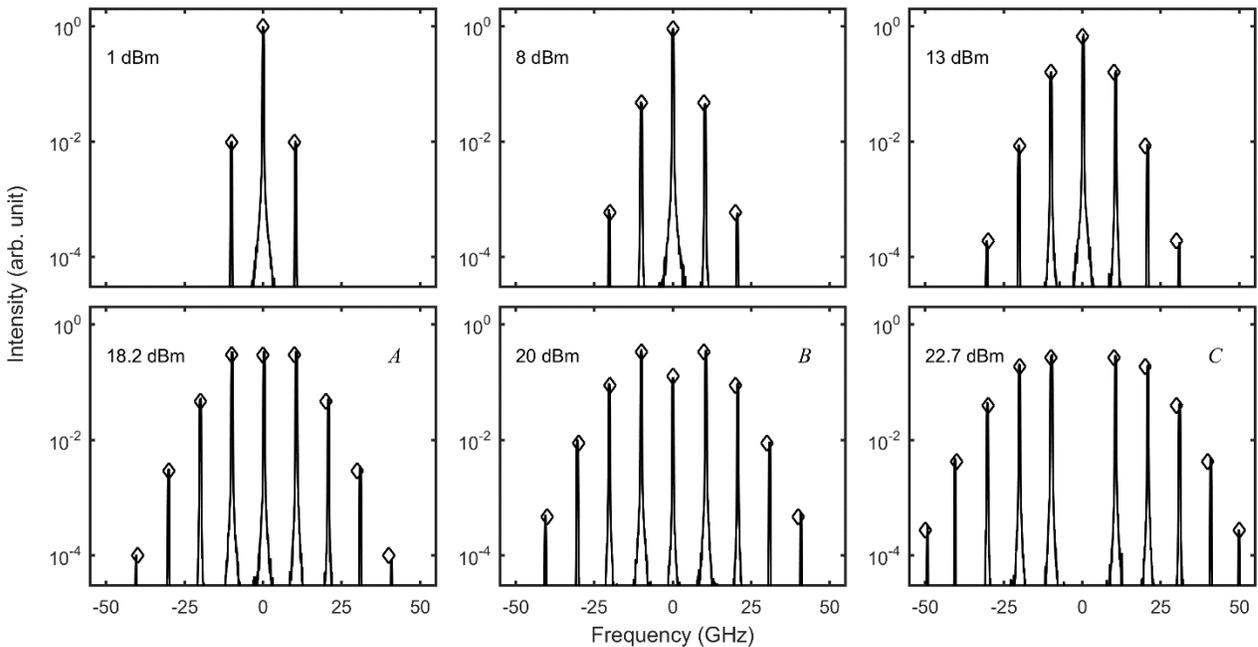

**Figure 5** – Optical spectra recorded on the high-resolution OSA for different average powers applied to the phase modulator. The diamonds represent the theoretical prediction from Eq.(6). The optical frequencies are relative to the central frequency $\omega_0$.

By changing the power of the electrical signal, i.e. $A_m$, we observe an increase in the extend of optical spectrum. Thus, the value of the measured spectral components varies and the students can see on the central component that this evolution is not monotonous. The link between these recordings and the theory is quite simple and can benefit of a Jacobi-Anger expansion [21, 22] of the modulated field:

$$\psi_m(t) = e^{i n \omega_0 t} \sum_{n=-\infty}^{\infty} i^n J_n(A_m) e^{i n \omega_m t}. \qquad (4)$$

The amplitude $\psi_n$ of the $n^{th}$ band is therefore analytically predicted from Bessel functions $J_n$ of the first kind and of (integer) order $n$ :

$$\psi_n \propto i^n J_n(A_m). \qquad (5)$$

Bessel functions, which are special mathematical functions [23, 24], are nowadays implemented in most of the scientific computation and programming software. We recently published an article interpreting this result qualitatively as a consequence of two-wave interference process [25]. Bessel functions appear in many different problems of wave propagation and static potentials, ranging from the study of the far-field diffraction through a circular aperture [26] to the vibration of a cantilever beam or a circular membrane [27]. One of our student commented that the spectral pattern reminds him of the diffraction he studied during an internship dealing with diffraction of a sinusoidal phase gratings. This is no coincidence since diffraction can be modelled using similar mathematical treatments and is consequently equivalent, leading to an exciting space/time duality [28].

The amplitude of Bessel functions is plotted in Fig. 6. The formula (5) highlights several properties illustrated in Fig. 6, which can be confirmed experimentally. First of all, the amplitude of the lines does not depend on the modulation frequency $\omega_m$ but only on the modulation contrast $A_m$. The evolution of Bessel functions is not monotonous. It is possible to have, for a certain modulation amplitude, $s_0 = s_1$ (point A, obtained for $A_m = 1.435$) or $s_2 = s_0$ (point B, $A_m = 1.841$) or even $s_0 = 0$ (point C, $A_m = 2.405$). Then, an important point to notice in Eq. (5) is the factor $i^n$ that leads to a phase offset of $\pi/2$ between two successive spectral components: the spectral components are not in phase so that the spectrum is not Fourier transform-limited.

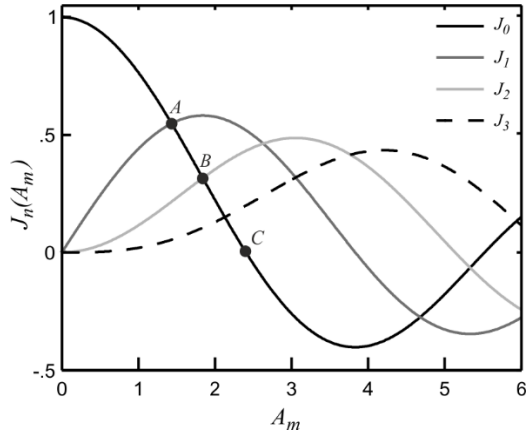

**Figure 6** – Evolution of the Bessel functions of the first kind with order from 0 to 3.

However, the optical spectrum made of line $S_n$ is the square modulus of the Fourier transform of the field:

$$S_n \propto |J_n(A_m)|^2. \qquad (6)$$

Therefore, it is not possible to visualize the relative sign change between the different components. By performing a series of measurements as a function of the electrical power, the students can plot the evolution of $S_n$ according to $V_m$ (Figure 7(a)). It should be noted that the students had to convert the modulation average powers into voltage by involving, in particular, the characteristic impedances of 50 Ω of the RF link. To compare the predictions of the Eq. (6) and the experimental measurements, it also remains to convert the value of the electrical modulation $V_m$ into optical modulation $A_m$. It is then interesting to look at the properties of point $A$, whose characteristic voltage can be measured with good accuracy. Indeed, for 2.51 V, we obtain a modulation amplitude such that $J_0=J_1$, i.e. $A_m = 1.435$. This therefore imposes the proportionality coefficient between the two quantities which is 0.57 rad/V. We can deduce an experimental value of the coefficient $V_\pi$ characteristic of the modulator: $V_\pi = 5.5$ V, in agreement with the manufacturer's technical documentation. Points $B$ or $C$ could also be used and lead to very similar values for $V_\pi$. Figure (7) shows the superposition of the experimental measurements and those expected from Eq. (6). The agreement obtained for the central and the 5 lateral sidebands is absolutely remarkable, even when displayed on a logarithmic scale. This proves the quality of the temporal phase modulation scheme as well as the quality of the spectral detection. From the experimental measurements, it can be observed that for an electrical power of 22.7 dBm, the central component is nearly completely vanished, with an extinction ratio close to 40 dB (see also the spectral record at this power on Fig. 5).

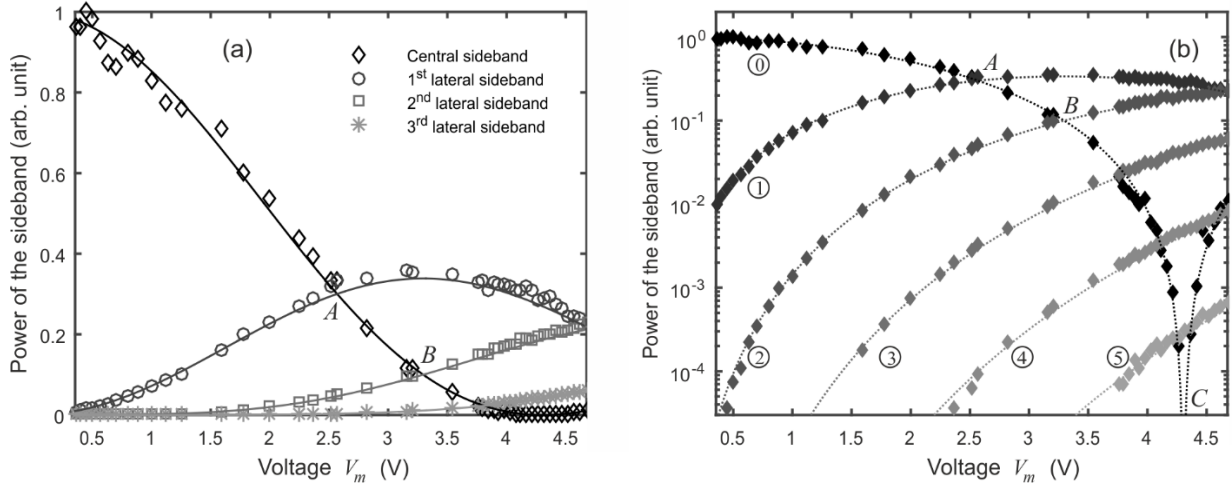

**Figure 7** – Evolution of the sidebands $S_n$ according to the maximum voltage $V_m$ applied to the phase modulator. **(a)** Results plotted on a linear scale for the central and the first three sidebands. **(b)** Results plotted on a logarithmic scale for the central and the 5 first lateral sidebands. The various solid and dashed lines represent the analytical predictions based on Bessel functions of the first kind (Eq. (6)).

### 3.3. Modification of the spectral properties.

In the final part of this lab session, our objective was to modify the spectral properties of the phase-modulated signal.

### 3.3.1. Use of a single-mode fiber spool.

The first method we exploit is extremely basic. We implement a standard optical fiber spool placed just after the modulator. Thanks to the power meter, students can first check that the attenuation in 2 km of fiber is very moderate, below 1 dB, including fiber connectors. The optical spectrum shows absolutely no change in the shape of the spectrum, keeping unaltered its frequency comb structure. Students therefore do not expect to observe any other changes, either on the ESA or on the oscilloscope. Indeed, for many of them, when nonlinear effects are neglected [29], the fiber stands as a communication channel that does not affect a continuous signal in intensity. For them, a phase change should not affect the measured temporal aspect. Nevertheless, their observations of the RF spectrum invalidate their first intuition. Indeed, as shown in Fig. 8(a), they observe a peak at the modulation frequency $\omega_m$. This very narrow peak is now perfectly discernible from the noise and is not due to electromagnetic compatibility concerns. They are also surprised to see a second peak at the frequency $2\omega_m$ (Fig. 8(b)). As the bandwidth of our ESA is limited to 26 GHz, the detection of additional harmonic peaks cannot be performed.

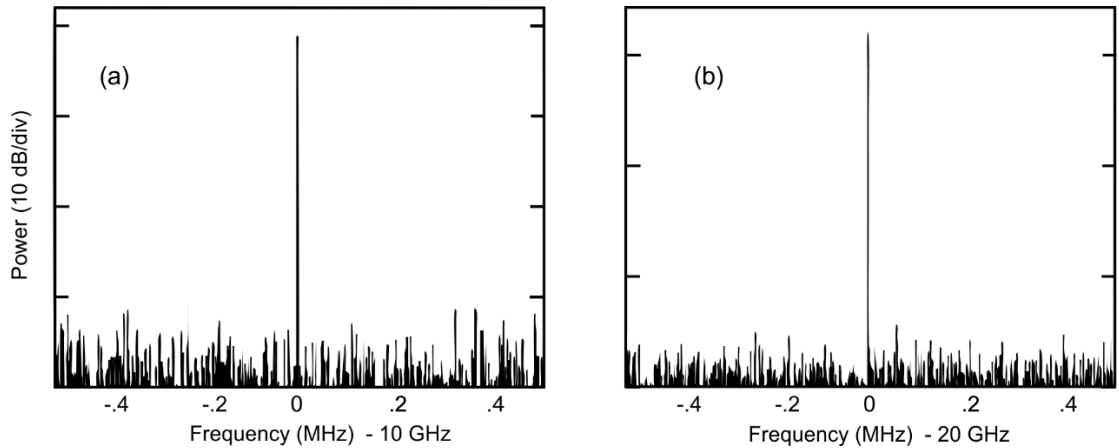

**Figure 8** – RF spectra recorded around $\omega_m$ and $2\,\omega_m$, panel **(a)** and **(b)** respectively.

If the RF spectrum is modified, it is because the temporal profile of the wave intensity is necessarily affected, one cannot go without the other. To visualize this change, we used fast oscilloscopes. Most students were then surprised to find that the initially continuous intensity profile has changed into a pulse profile lying over a continuous background (Figure 9). The use of an electrical oscilloscope provides a full-width at half maximum duration around 10 ps that is confirmed by the optical sampling oscilloscope. The repetition rate is dictated by the modulation frequency so that the frequency of the optical train is directly adjustable by the electrical signal generator. This reshaping of the temporal intensity content highlights the different but symmetrical roles played by the temporal and spectral phases. If modifying the temporal phase impacts the spectral intensity profile without affecting the time intensity profile, the opposite happens when the spectral phase is modified.

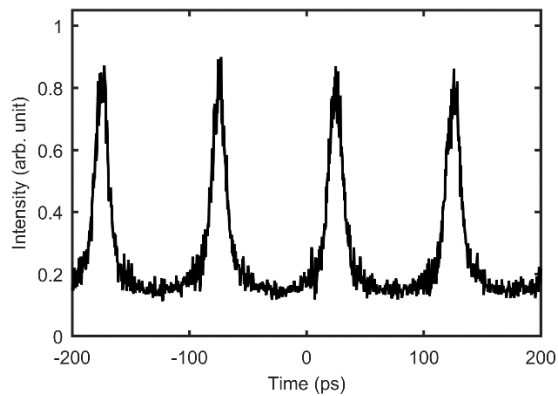

**Figure 9** – Temporal intensity profile after propagation in a 2-km long optical fiber. The amplitude of the initial phase modulation $A_m$ is around 2 rad.

For a qualitative understanding of the process stimulating the emergence of those temporal structures, it should be remembered that temporal phase modulation induces a modulation of the pulse frequency. Thus, as can be seen in Fig. 10, the continuous wave will see its instantaneous frequency, or chirp, $\delta\omega(t)$ vary over time with the presence of areas with higher frequencies than the carrier and others with lower frequencies. The parabolic phase applied in the spectral domain can be actually interpreted as a delay applied to the different spectral components of the modulated signal: the low frequency components are delayed with respect to the central frequency whereas the high frequency components are advanced. By applying this spectral parabolic phase shift, one reduces the impact of the initial $\pi/2$ spectral phase offset between the various components that we previously mentioned (see Eq. (5)). Therefore, the parabolic phase leads to a convergence of energy in certain time slots, creating a pattern of pulses repeating at the frequency of the initial modulation. Such a feature has been exploited in the context of optical retiming of a stream of data impaired by temporal jitter [30]. As can be seen in Fig. 10(b), the numerical results are in very good agreement with the experimental recordings. Shorter structures with an improved peak power can be achieved using an optimized length of fiber (see dotted-dashed line obtained for a simulated length of 5 km). The obvious application of this process is in the field of high-speed optical pulse generation [31, 32] where the combination of time phase modulation followed by parabolic spectral phase modulation generates an attractive all-optical solution for high repetition rate pulse train generation. It should be also noted that the analogy between the temporal and spatial domains [33, 34] makes it possible to reinterpret this dynamic and to associate to this sequence the case of a lenticular lens [35] that can be used for high speed optical sampling. For these two types of applications, our team had the opportunity to publish various contributions in international journals, which allowed us to make a direct link between the proposed practical work and our academic research.

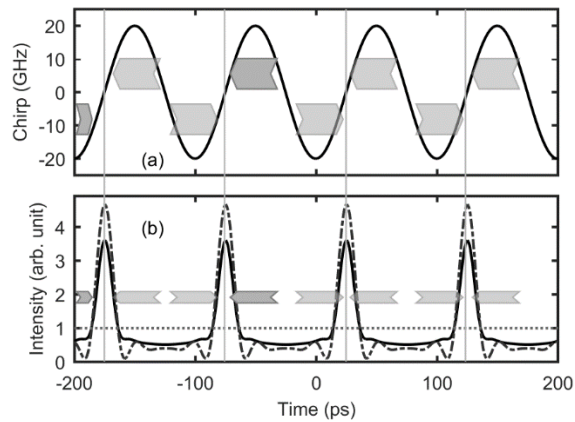

**Figure 10** – Numerical simulations of the temporal waveform properties: **(a)** Instantaneous frequency obtained after phase modulation. (b) Intensity profile monitored directly after temporal sinusoidal phase modulation (dash line), after propagation in a fiber of 2 km (solid black line) and of 5 km (dash dotted line). The various grey arrows indicate the impact of the fiber dispersion.

### 3.3.2. Use of a spectral phase shaper.

We then led the students to use a spectral shaper. At first, we only used phase shaping features. In particular, the shaper allows the inclusion of an additional spectral parabolic phase [36]. By playing on the parabolic phase level, the students were able to find the results previously obtained with a fiber. The students were also able to check that the sign of dispersion for this application does not matter. However, a limit appears for strong phases for which attenuation of certain spectral components may be visible.

The use of spectral shaper as dispersion emulator [36] confirms that a fiber induces nothing but a parabolic spectral phase. The fiber retains some advantages such as its low cost and its very low level of attenuation, much lower than that introduced by linear shaping (4 dB of loss related to the component). However, the shaper has the advantage of flexibility. It allows the value of the dispersion to be precisely adjusted to obtain the shortest pulse train. In more advanced architecture, fine tuning of the spectral phase of a limited set of spectral lines can also enable generation of parabolic, triangular or rectangular intensity profiles [37]. This has stimulated strong practical interest for applications in the field of microwave photonics where manipulating light can help to generate electrical waveforms that would be not possible with the current bandwidth limitations of electrical systems [38].

### 3.3.3. Use of a spectral intensity shaper.

Finally, the students tested a spectral intensity shaping, i.e. applying a bandpass optical filter to the signal by means of the previous waveshaper. We asked them to compute a Gaussian bandpass filter with a full width at half maximum of 25 GHz, offset by 25 GHz from the center frequency. The programming is very simple and the students were able to visualize directly on the OSA the result of this filtering which eliminates a number of components and reduces the amplitude of other components. We asked them to experimentally check that the spectral filter corresponded to our expectations. This question was not so trivial for them. Indeed, the simplest approach is to basically compute the ratio between the output signal and the input signal. But the comb nature of the spectrum complexifies slightly the analysis, particularly because of the detection noise between the spectral lines, which disturbs the ratio. Students must only take into account the level of the narrow spectral lines and then use a fitting by a Gaussian shape to verify that the experimental points they have, are in agreement with the target.

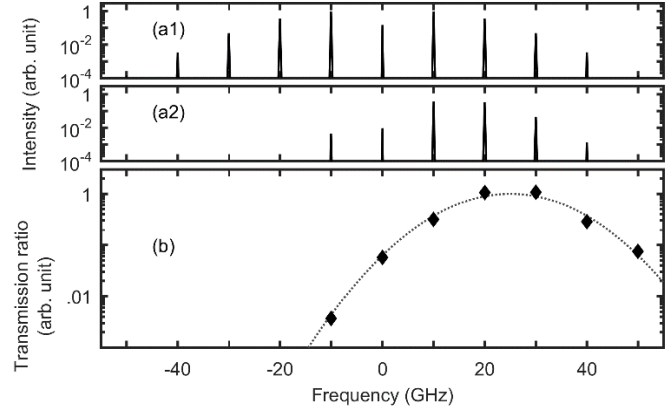

**Figure 11 –(a)** Optical spectrum before (a1) and after (a2) the optical bandpass filter. (b) Ratio between the input and output levels of the various spectral lines (diamonds) and the corresponding fit by a Gaussian waveform. Result are plotted on a logarithmic scale and the optical frequency are provided with respect to the carrier frequency $\omega_0$. Results are here obtained for a power of 21 dBm leading to $A_m = 2$ rad. Optical frequencies are relative to the optical carrier $\omega_0$.

The students also noted that this adjustable filter did not allow the precise generation of filtering profiles with a width below 10 GHz without adding a noticeable level of losses. We asked our students regarding the temporal intensity profile that would be observed if our filter were thin enough to isolate a single spectral line shifted by $\omega_m$, 2 $\omega_m$ or 3 $\omega_m$. Many students spontaneously answered a sinusoidal evolution of the intensity profile at the angular frequency $\omega_m$, 2 $\omega_m$ or 3 $\omega_m$. The correct answer is rather different and the students are confused once again between the frequencies affecting the carrier and the envelope. Indeed, by isolating a single spectral line, we get a continuous wave, whose carrier frequency has been frequency offset. This opens very exciting perspective for wavelength multiplexed communications: for example, isolating the spectral lines of a supercontinuum sources has enabled the generation of more than a thousand independent transmission channels [39]. Similar motivation currently drives the field of optical microresonators [40]. In both cases, the initial comb is not generated thanks to a phase modulator, but results from the self-phase modulation that affects any high-power pulse propagation in a nonlinear optical waveguide [41].

When monitoring the average power on the power meter, a number of students were surprised by the rather high level of losses observed following the intensity filtering: more than 10 dB had simply disappeared. After reflection on their part, they made the link with the suppression of spectral components which is accompanied by a loss of energy. Unlike advanced non-linear shaping techniques [42], linear intensity shaping is a dissipative operation that does not allow any redistribution of energy between the different components. This significant loss of energy makes the detection on the oscilloscope quite noisy. We then include in the assembly an Erbium doped fiber amplifier that allows to regain the lost energy. Nevertheless, students notice an increased in the noise level on the spectrum, with amplification accompanied by noise emission. A discussion then allows them to understand that it is preferable to place the amplifier before the spectral shaper because the latter will help to eliminate the noise brought in.

The last part of this work deals with the experimental observation of the resulting temporal signal on the oscilloscope. The students then observe a pulse train at a frequency corresponding to the initial phase modulation. By tuning the wavelength offset of the filter, they can see that the further away the filter is from the central line, the more the generated pulses are isolated and visible (Fig. 12(a)). Students find strong marks of this periodicity on the RF spectrum with significant lines at 10 GHz and 20 GHz, confirming that the signal is not a pure sinusoid. A frequency-offset optical bandpass filtering is therefore a second convenient solution to convert the phase modulation into a pulse train. This solution is also used in research [43] as well as in telecommunications [44-46]. Again, it should be noted that such a behavior does not appear natural for students that are used to the fact that when the spectral width is reduced, the duration should increase. Here, it is the opposite, the limitation of the spectral range has made it possible to create shorter structures. Once again, it is the notion of instantaneous frequency $\delta\omega$ that will allow us to understand this phenomenon qualitatively. It should be recalled that in the previous method, the parabolic spectral phase made it possible to converge energy at specific times without loss of energy. With wavelength-shifted filtering, the approach is different and inherently dissipative. As illustrated in Fig. 12(b), the bandpass filter only keeps the frequencies corresponding to one extrema and strongly attenuates the other components. Thus, the energy contained at times when the instantaneous frequency is different from the filter central frequency is vanished, leading to light holes and consequently the emergence of light pulses. Simple numerical simulations, displayed in Fig. 2b, can reproduce these experimental results.

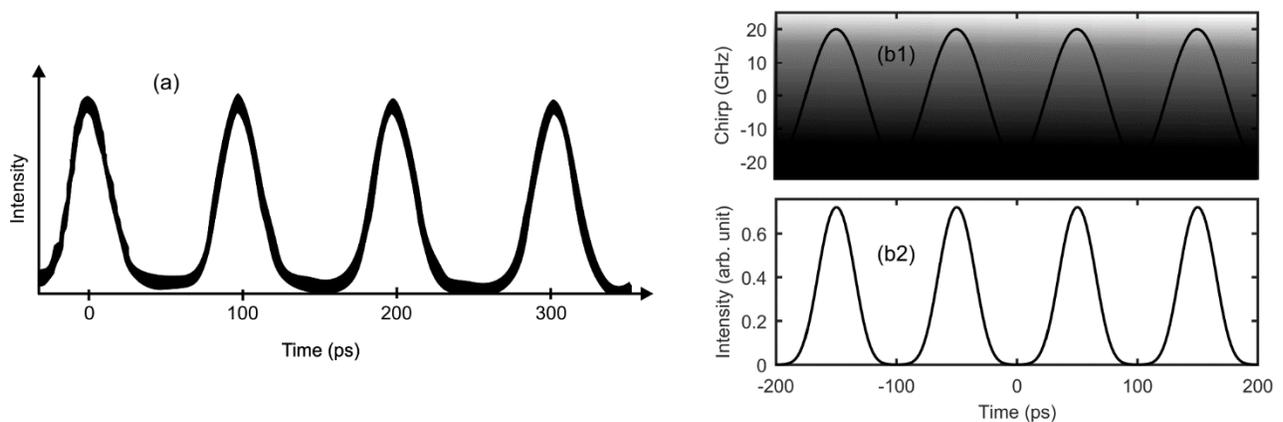

**Figure 12** –**(a)** Experimental recording of the pulse train obtained after frequency offset optical bandpass filtering. (b) Numerical simulations of the instantaneous frequency (b1) and of the temporal intensity profile (b2). The shaded area represents the instantaneous frequency that are suppressed by the optical bandpass filter. Results are here obtained for a RF power of 21 dBm leading to $A_m = 2$ rad.

## 4. Assessment of the students results

This laboratory session follows an 8-hour lecture detailing the various parameters characterizing the temporal, spectral and spatial properties of light and introducing -all the theoretical concepts required for a good understanding of the session which lasted half a day (3 hours). A detailed description of the work expected during this lab session and containing all the technical information of the experimental devices was made available well in advance to the students. The experiments took place in the permanent presence of a teacher or researcher. Such availability is justified on the one hand by the high cost and the specificity of the equipment made available to the students and on the other hand by our will to establish a live interaction throughout the lab session regarding the various experimental observations. We noted the need for numerous reminders to correct some of the erroneous pictures that students could have in mind for example about the optical phase or, more generally, about the spectra they recorded. The students worked in pairs or trinomials and were evaluated during this session on their personal involvement (15% of the final mark), our goal being to avoid any passivity of the students. Since we wanted to stimulate exchanges, we did not take into account the errors made or the questions.

Following the lab session, the students had to write a detailed report of their work (one report per pair or trinomial). They had several weeks to complete this report and we encouraged them to exchange between different groups as needed. They have also been strongly encouraged to use software such as Matlab® or its clones to reproduce numerically and validate their experimental observations. Voluntarily, no ready-to-use code was provided to them, their experience gained in the other teaching modules being sufficient to allow them to be autonomous on this numerical task.

The next step was to simulate for our master students the reviewing process as known in peer-reviewed journals. Thus, the various reports were exchanged between the teams and everyone had to assess the strengths and weaknesses of their classmates. The teams then had the opportunity to respond to these comments and modify their work accordingly. The reports have been evaluated by the teaching team at both stage, i.e. before and after the reviewing process (30% and 15 % of the final mark respectively). The relevancy of the review of the other team has also been evaluated (15% of the final mark). Regarding the assessment of final report, the evaluation has been focused on how accurately the different comments raised by the review have been answered and how it has translated into improvements of the manuscript.

The final reports corrected by the teachers were given back to the students before the last stage of the evaluation process, consisting of a short oral interview (about ten minutes) to determine the overall understanding of the main theoretical and experimental concepts discussed in the framework of this lab session. This last step was carried out individually and was included as 25% of the mark.

## 5. Conclusions

This lab session carried out over 4 years, from 2012 to 2016, with a total of about forty students. The feedback was very positive, both from the point of view of the master students, who were delighted to handle state-of-the-art equipment on a concrete problem, and from the point of view of the supervisors, who strongly appreciated the personal investment shown by a few teams. The sessions provided a real dialogue around the importance of the optical phase, telecommunications and the concept of optical and electrical spectra, correcting a number of misconceptions that students had in mind. This practical work clearly helped the students to better understand the usefulness and complementarity of the different analyses that can be carried out on a signal and the nuances between electrical and optical spectra. Thus, a contradicted idea is that the electrical spectrum does not correspond to a simple enlargement of the optical spectrum intended only to visualize the lowest frequencies with greater precision. We have shown to the students that even a modulation of phase as simple as a sinusoidal waveform can already be relevant for many modern applications of fiber optics [28, 30-32, 35, 43, 47, 48]. The students have also found the discussion helpful to understand nonlinear effects, and more specifically why the self-phase modulation induced by high power pulse leads to a spectral broadening [29, 47, 49]. We have found that the practical examples displayed directly on the measuring devices can be a very valuable complement to the numerical simulation tool by reducing the potential passivity of some students. Finally, these lab sessions have also enabled us to discuss academic research subject that are currently under investigation in different groups. It shows that the legacy of analysis proposed by Joseph Fourier [50] (2018 marks the 250th anniversary of his birth in our region) remains of a tool of the most crucial importance in modern applications such as high-speed transmissions .

**Acknowledgement**

The authors would like to acknowledge the financial support from the Région Bourgogne Franche-Comté (Pari Photcom), the Agence Nationale de la Recherche (Labex Action ANR-11-LABX-01-01) and the Institut Universitaire de France.